# Soliton trains in photonic lattices

Yaroslav V. Kartashov, Victor A. Vysloukh, Lluis Torner

*ICFO-Institut de Ciencies Fotoniques, and Department of Signal Theory and Communications, Universitat Politecnica de Catalunya, 08034 Barcelona, Spain*

We address the formation and propagation of multi-spot soliton packets in saturable Kerr nonlinear media with an imprinted harmonic transverse modulation of the refractive index. We show that, in sharp contrast to homogeneous media where stable multi-peaked solitons do not exist, the photonic lattices support stable higher-order structures in the form of *soliton packets,* or *soliton trains*. Intuitively, such trains can be viewed as made of several lowest order solitons bound together with appropriate relative phases and their existence as stable objects puts forward the concept of compact manipulation of several solitons as a single entity.

***OCIS codes:*** *(190.0190) Nonlinear optics; (190.5530) Pulse propagation and solitons*

Wave propagation in nonlinear media whose properties vary periodically along the transverse direction exhibits a wealth of new features including a possibility of lattice soliton formation. In strongly guiding and evanescently coupled nonlinear waveguide arrays the formation of discrete solitons [1,2] is possible, that have attracted great attention because of their potential for all-optical switching and routing. In this context the intermediate regime constituted by continuous nonlinear media with an imprinted transverse modulation of the refractive index offers a variety of new opportunities. The concept behind such regime might be termed *tunable discreteness*, with the strength of modulation being the parameter that tunes the system properties from those expected for systems modeled by continuous equations, to those exhibited by systems modeled with discrete evolution equations [3,4]. Recently lattice solitons were studied in arrays of optically induced waveguides in photorefractive crystals [5-13]. Basic properties of lowest order lattice solitons are well understood. Some examples of higher-order multi-humped lattice solitons in Bose-Einstein condensates or Kerr optical media were found in Refs [14,15]. However, the existence and rigorous stability analysis of complex lattice soliton states, in the form of soliton packets, in saturable media is an open problem.

In this paper we report the results of a detailed analysis of the properties and dynamical stability of different families of one-dimensional lattice solitons in both focusing and defocusing saturable optical medium with harmonic transverse modulation of linear refractive index. Stability analysis reveals the existence of stability regions for even solitons that broaden with increase of degree of saturation of nonlinear response. We discovered existence of stable multi-soliton structures in the form of *solitons packets*, or *soliton trains*, that are analogous to twisted strongly localized modes of discrete systems. Intuitively, such packets can be built from an arbitrary number of lowest order odd solitons, or *soliton bits,* with appropriately engineered phases. No examples of such structures are known in uniform media, thus the photonic lattice uniquely affords the results reported here.

We consider light beam propagation along $z$ axis of a saturable medium with a periodic modulation of the linear refractive index in $x$ direction. We use a mathematical model set by the inhomogeneous saturable nonlinear Schrödinger equation:

$$i\frac{\partial q}{\partial \xi} = -\frac{1}{2}\frac{\partial^2 q}{\partial \eta^2} + \frac{\sigma q\left|q\right|^2}{1+S\left|q\right|^2} - pR(\eta)q, \tag{1}$$

where the transverse $\eta$ and the longitudinal $\xi$ coordinates are scaled to the characteristic beam width and diffraction length, respectively; $S$ is the saturation parameter; $\sigma = -1\,(+1)$ for focusing (defocusing) media; $p$ is the guiding parameter proportional to the linear refractive index modulation depth; $R(\eta) = \cos(2\pi\eta/T)$ describes the refractive index profile, and $T$ is the modulation period. The depth of linear refractive index modulation is assumed to be comparable with nonlinear contribution to refractive index. Note that Eq. (1) admits several conserved quantities including the energy flow

$$U = \int_{-\infty}^{\infty} \left|q\right|^2 d\eta. \tag{2}$$

We stress that the concept put forward here is expected to hold for systems described by other models, like photonic lattices in photorefractive media, and in Bose-Einstein condensates.

Stationary solutions of Eq. (1) have the form $q(\xi,\eta)=w(\eta)\exp(ib\xi)$, where $w(\eta)$ is a real function and $b$ is the real propagation constant. Lattice soliton families are defined by the propagation constant $b$, modulation period $T$, guiding and saturation parameters $p$, $S$. Since scaling transformation $q(\eta,\xi,S,p)\rightarrow\chi q(\chi\eta,\chi^2\xi,\chi^{-2}S,\chi^2 p)$ can be used to obtain various families of lattice solitons from a given one, we selected the transverse scale in such way that modulation period $T=\pi/2$, and vary $b$, $S$, and $p$. Families of stationary solutions were obtained by solving Eq. (1) with a relaxation algorithm. Standard linearization procedure was used to analyze soliton stability and obtain perturbation growth rates $\delta$.

For focusing nonlinearity optical lattices support two types of lowest order solitons: *odd* an *even* (Fig. 1). Maximum of odd soliton coincides with the maximum of $R(\eta)$ (Fig. 1(c)), while even one is centered between neighboring maximums of $R(\eta)$ (Fig. 1(d)). Dispersion curves $U(b)$ for odd and even solitons almost coincide (Fig. 1(a)). There are lower and upper cutoffs on propagation constant for both odd and even solitons. Lattice solitons transform into Bloch waves in both cutoffs. Lower cutoff $b_{\text{low}}$ is the monotonically growing function of $p$ (Fig. 1(b)), while upper cutoff is given by $b_{\text{upp}}=b_{\text{low}}-\sigma/S$. The segment of existence on $b$ shrinks with growth of saturation parameter $S$ for both odd and even lattice solitons.

Linear stability analysis revealed existence of instability areas for odd lattice solitons associated with real perturbation growth rates near the upper cutoff (Fig. 2(a)). The width of instability areas decreases as $b\rightarrow b_{\text{upp}}$ and as $S\rightarrow 0$. Growth of guiding parameter $p$ results in decrease of instability areas width. We revealed that nonlinearity saturation stabilizes even lattice solitons that are unstable in discrete systems [1,2]. Notice that instability areas for odd solitons and stability areas for even ones coincide (Figs. 2(a) and 2(b)). Stable propagation of odd and even solitons perturbed by white noise is illustrated in Figs. 2(c) and 2(d).

Next we address properties of *twisted* lattice solitons (Fig. 3). In such solutions, the function $w(\eta)$ changes its sign between neighboring maxima of the lattice (Fig. 3(c)), so intuitively they can be viewed as nonlinear superposition of several odd solitons with appropriately engineered phases. Fig. 3(a) shows the corresponding dispersion curves. The range of existence on propagation constant decreases with growth of twisted soliton order. Lower and upper cutoffs for lowest order (or first) twisted soliton are shown in Fig. 3(b). The area of existence shrinks with growth of saturation parameter and broadens with increase of guiding parameter. Stability analysis revealed that first twisted soliton becomes stable over certain threshold value of energy flow (Fig. 3(d)). Notice that the structure of shaded area in Fig. 3(d) is complex, with separate “stability windows”, but we do not show them here. Also there exist instability band near upper cutoff that is narrow and not seen in the plot.

The central result of this paper is that *stabilization takes place for twisted solitons of arbitrary higher order,* above a certain threshold energy level. Figures 3(e) and 3(f) show the profile and the stable propagation of higher order twisted soliton perturbed by broadband noise. Since twisted solitons can be intuitively seen as nonlinear superposition of several odd solitons with opposite phases, this result implies that “multi-soliton trains” can be built out of several individual solitons. We found that the threshold energy flow for stabilization slowly increases with growth of soliton order (i.e., with the number of spots in the soliton packet). On physical grounds, the stabilization of such higher-order soliton trains can be understood as follows. Because of the alternating phases of neighboring lowest-order spots involved into the packets, the interaction forces acting between them are repulsive, so such higher-order soliton complexes cannot exist as the stable objects in uniform media. The periodic potential set by the lattices compensates such repulsive forces, thus enabling stable propagation.

As pointed out above, the lattice strength is one of the key parameters for control of soliton properties. For example, in the presence of the periodic refractive index modulation, at a given energy level, the field amplitude necessary for soliton-like propagation amounts to lower values than that in homogeneous focusing media. Therefore, at fixed energy level solitons in lattices with higher values of $p$ feature broader, more extended shape. This in turn reduces nonlinearity saturation at the soliton peak and thus broadens the energy interval where single solitons and soliton trains exist (inset in Fig. 3(b)). Thus, the energy level necessary to form long soliton trains can be substantially reduced by increase of the lattice depth.

Another example of the importance of the lattice strength for control of soliton features is illustrated by removing the lattice and letting the train to split or unpack. Such splitting produces sets of diverging solitons with controllable escape angles (Fig. 4(a) and 4(b)). The higher is the number of solitons in the train the faster is the train splitting. We calculated escape angles for solitons that appear upon decay of the soliton trains and here we show results for the simplest train. Increasing the overall energy flow of the train (Fig. 4(c)) causes the escape angle to decrease first, because of decrease of tails overlap for neighboring spots in the train, but then angle start to grow, because nonlinearity saturation tend to increase widths of spots located in neighboring lattice sites and, hence, the interaction forces between them. Increase of the transverse extent of spots forming the train with growth of $p$ at fixed overall energy flow (Fig. 4(d)) also causes faster train splitting. This is one of manifestation of unique features afforded by tunability of the lattice depth that is not possible in fully discrete systems. Notice that the splitting of multi-humped solitons into sets of diverging fundamental solitons mediated by instability was also reported in uniform quadratic media [16].

Lattices with defocusing nonlinearity also support localized self-sustained light beams (Fig. 5). They are typically wider than solitons in focusing lattice and have several amplitude oscillations (Fig. 5(c)). Intuitively they can be viewed as a result of coupling of fundamental modes guided by separate lattice sites. Inside each site diffraction and defocusing nonlinearity are compensated by linear refraction. Fundamental modes forming odd soliton are out-of-phase, while for even solitons they are in-phase (Fig. 5(c) and 5(e)). Dispersion curve $U(b)$ is shown in Fig. 5(a) for odd soliton. Lower and upper cutoffs are presented in Fig. 5(b). We have found that for odd solitons the multiple instability bands appear near lower cutoff, while close to upper cutoff odd solitons are stable. Propagation of perturbed odd soliton is shown in Fig. 5(d). The important result is that defocusing lattices also support stable even solitons of different orders. Such solitons conserve their input structure up to the several thousands of diffraction lengths in the presence of white noise (Fig. 5(f)). Notice that twisted solitons in defocusing lattices are exponentially unstable in the entire domain of their existence.

In conclusion, we analyzed properties of lattice solitons in Kerr-type saturable medium with harmonic transverse modulation of the linear refractive index, and revealed that besides the simplest stable odd solitons, such lattices support stable even and twisted solitons that form the trains. Such multi-spot soliton states can be viewed as nonlinear superposition of several solitons stuck together into stable packets, and thus open the possibility of the compact manipulation of several solitons as a single state.

This work has been partially supported by the Generalitat de Catalunya and by the Spanish Government through grant BFM2002-2861.

## Figure captions

Figure 1. (a) Energy flow versus propagation constant for odd and even lattice solitons at $p = 3$, $S = 2$. (b) Lower cut-off for odd and even solitons versus guiding parameter at $S = 0.5$. Inset shows lower and upper cut-offs for odd and even solitons versus saturation parameter at $p = 3$. Profiles of odd (c) and even (d) lattice solitons at $p = 3$, $S = 2$, $b = 0.8$. Focusing medium $\sigma = -1$.

Figure 2. Areas of stability and instability (shaded) for odd (a) and even (b) solitons on the $(b, S)$ plane at $p = 3$. Stable propagation of odd (c) and even (d) solitons with $b = 0.8$ at $p = 3$, $S = 2$, perturbed with white noise with variance $\sigma^2_{\text{noise}} = 0.02$. Focusing medium $\sigma = -1$.

Figure 3. (a) Energy flow versus propagation constant for twisted lattice solitons of first three orders at $p = 3$, $S = 0.5$. (b) Lower and upper cut-offs for first twisted soliton versus saturation parameter at $p = 3$. Inset shows cut-offs versus guiding parameter at $S = 0.4$. (c) Profile of first twisted soliton at $p = 3$, $S = 0.4$, $b = 1$. (d) Areas of stability and instability (shaded) for first twisted soliton on the $(b, S)$ plane at $p = 3$. (e) Profile of third twisted soliton at $p = 3$, $S = 0.25$, $b = 2$, and (f) its stable propagation in the presence of white input noise with variance $\sigma^2_{\text{noise}} = 0.02$. Focusing medium $\sigma = -1$.

Figure 4. Decay of first (a) and second (b) twisted solitons with $U = 50$ at the boundary of uniform medium and medium with periodic modulation of refractive index at $p = 3$. Horizontal dashed line shows boundary between two media. Escape angle for solitons that appear upon decay of first twisted soliton versus energy flow at $p = 3$ (c), and versus guiding parameter at $U = 20$ (d). Focusing medium $\sigma = -1$. Saturation parameter $S = 0.05$.

Figure 5. (a) Energy flow versus propagation constant for odd lattice soliton at $p = 3$, $S = 1.5$. (b) Lower and upper cut-offs for odd soliton versus saturation parameter at $p = 3$. Inset shows lower and upper cut-offs versus guiding parameter at $S = 0.5$. (c) Profile of odd lattice soliton at $p = 3$, $S = 1.5$, $b = -0.45$, and (d) its stable propagation in the presence of white noise. (e) Profile of even soliton at $p = 3$, $S = 0.4$, $b = -1.5$, and (f) its stable propagation in the presence of white noise. Noise variance in (d) and (f) $\sigma^2_{\text{noise}} = 0.02$. Defocusing medium $\sigma = 1$.

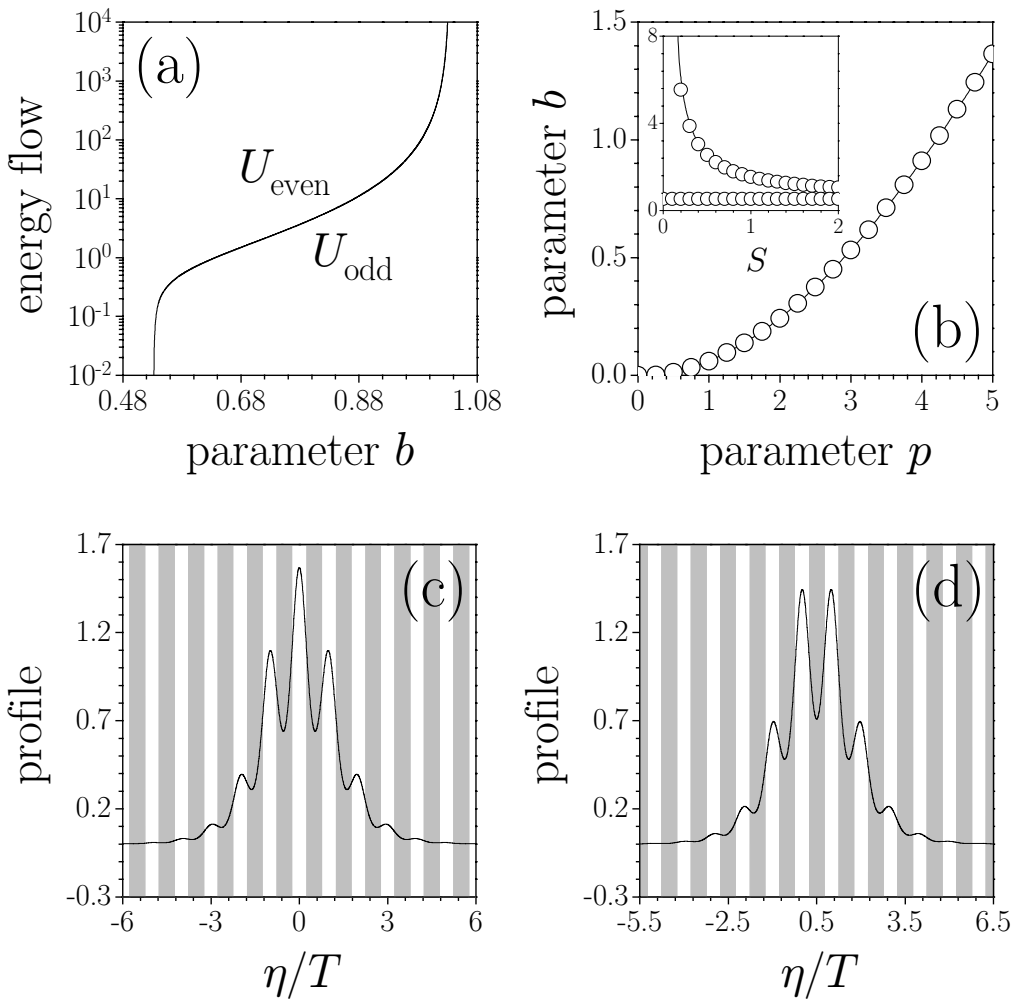

Figure 1. (a) Energy flow versus propagation constant for odd and even lattice solitons at $p = 3$, $S = 2$. (b) Lower cut-off for odd and even solitons versus guiding parameter at $S = 0.5$. Inset shows lower and upper cut-offs for odd and even solitons versus saturation parameter at $p = 3$. Profiles of odd (c) and even (d) lattice solitons at $p = 3$, $S = 2$, $b = 0.8$. Focusing medium $\sigma = -1$.

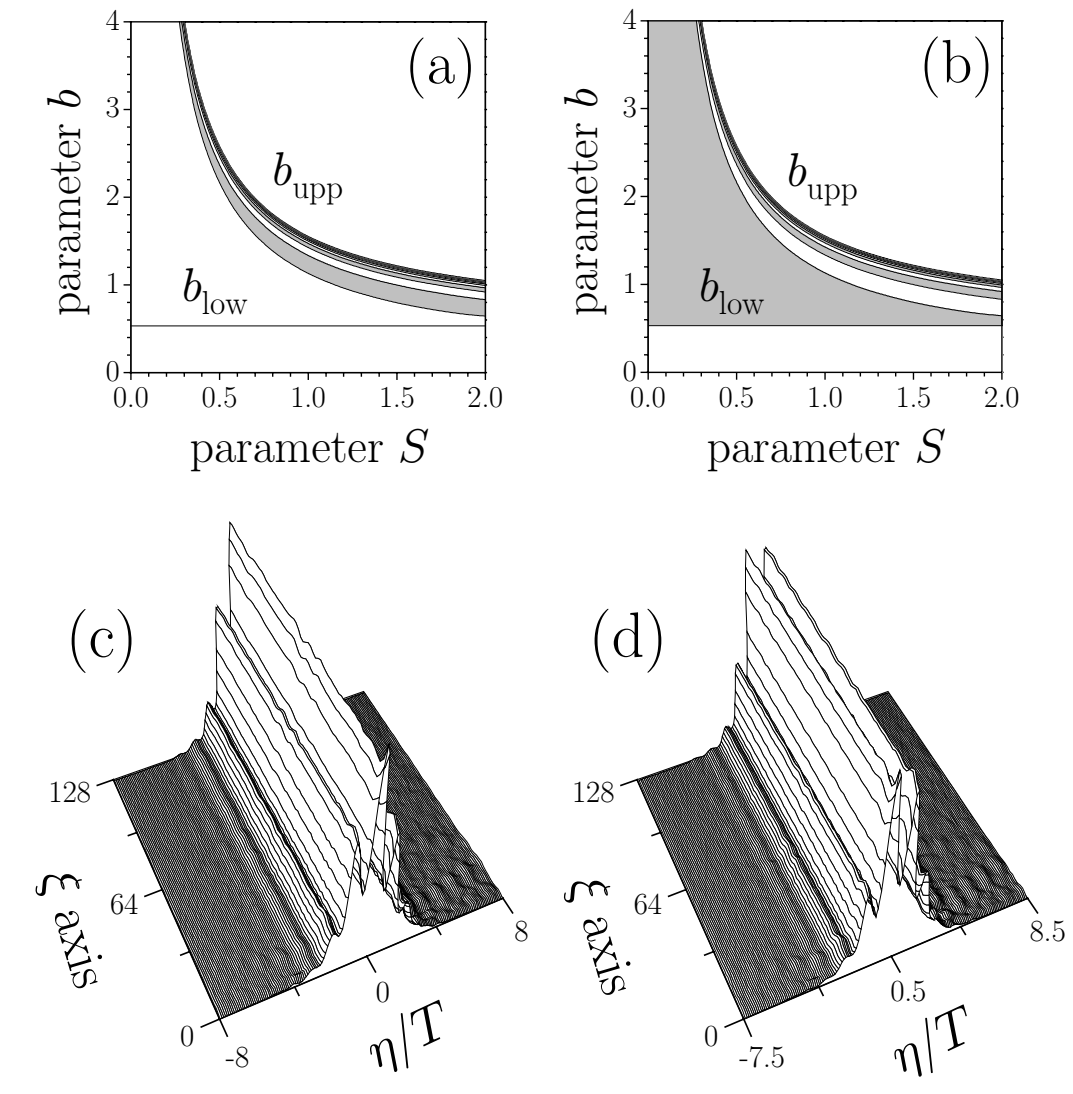

Figure 2. Areas of stability and instability (shaded) for odd (a) and even (b) solitons on the $(b,S)$ plane at $p = 3$. Stable propagation of odd (c) and even (d) solitons with $b = 0.8$ at $p = 3$, $S = 2$, perturbed with white noise with variance $\sigma^2_{\text{noise}} = 0.02$. Focusing medium $\sigma = -1$.

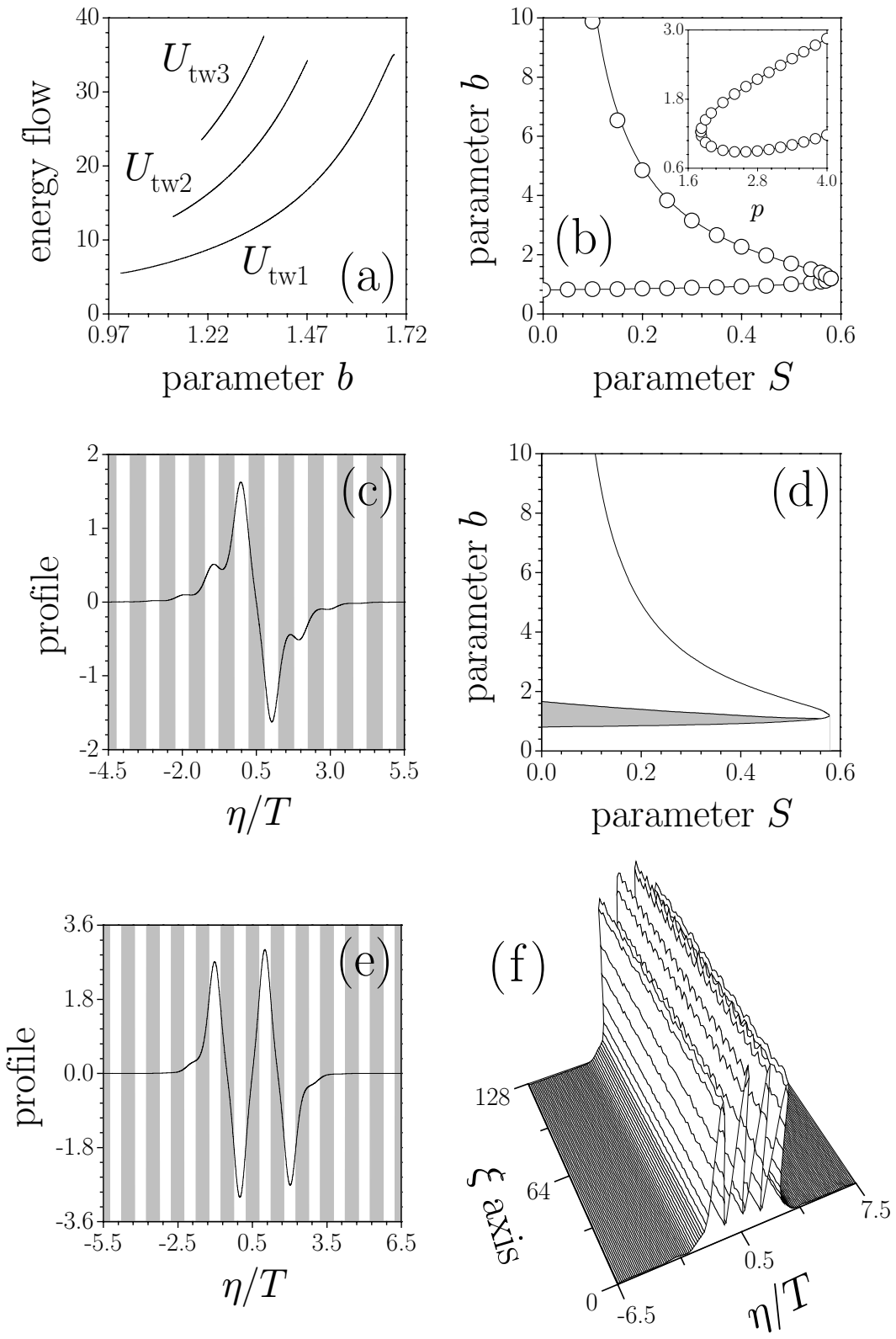

Figure 3. (a) Energy flow versus propagation constant for twisted lattice solitons of first three orders at $p = 3$, $S = 0.5$. (b) Lower and upper cut-offs for first twisted soliton versus saturation parameter at $p = 3$. Inset shows cut-offs versus guiding parameter at $S = 0.4$. (c) Profile of first twisted soliton at $p = 3$, $S = 0.4$, $b = 1$. (d) Areas of stability and instability (shaded) for first twisted soliton on the $(b,S)$ plane at $p = 3$. (e) Profile of third twisted soliton at $p = 3$, $S = 0.25$, $b = 2$, and (f) its stable propagation in the presence of white input noise with variance $\sigma^2_{\text{noise}} = 0.02$. Focusing medium $\sigma = -1$.

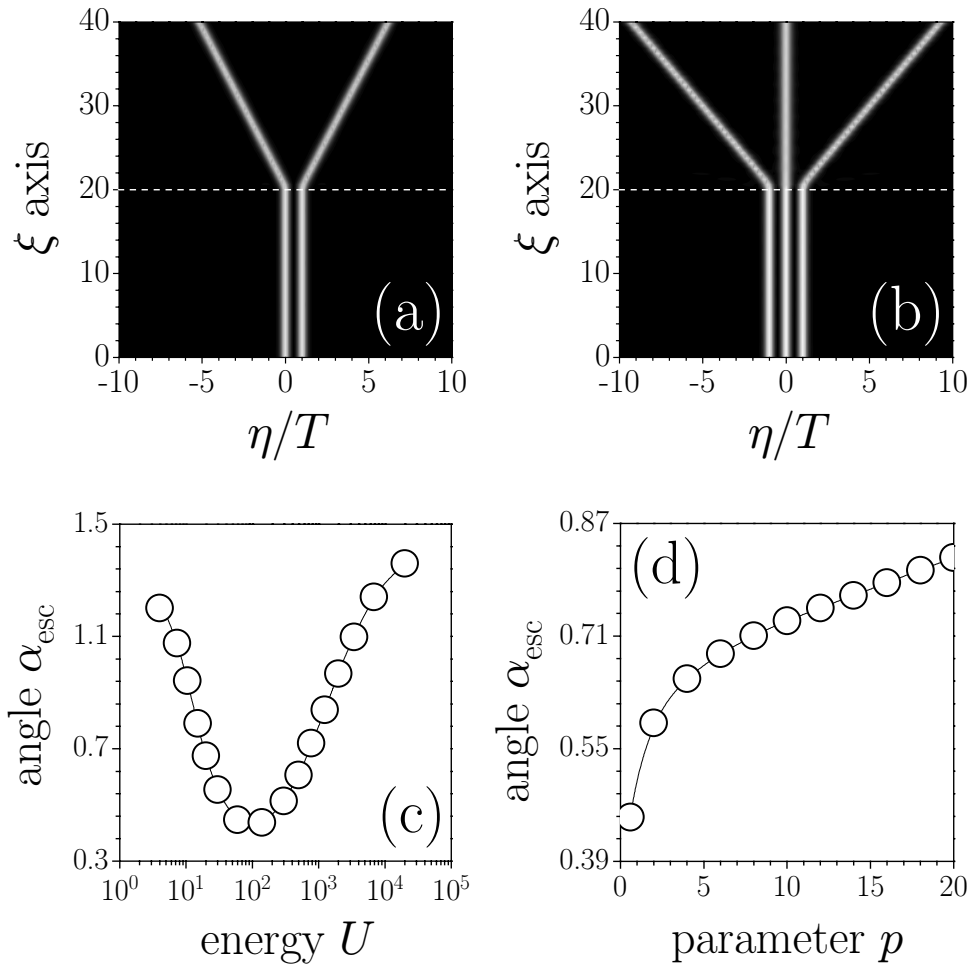

Figure 4. Decay of first (a) and second (b) twisted solitons with $U = 50$ at the boundary of uniform medium and medium with periodic modulation of refractive index at $p = 3$. Horizontal dashed line shows boundary between two media. Escape angle for solitons that appear upon decay of first twisted soliton versus energy flow at $p = 3$ (c), and versus guiding parameter at $U = 20$ (d). Focusing medium $\sigma = -1$. Saturation parameter $S = 0.05$.

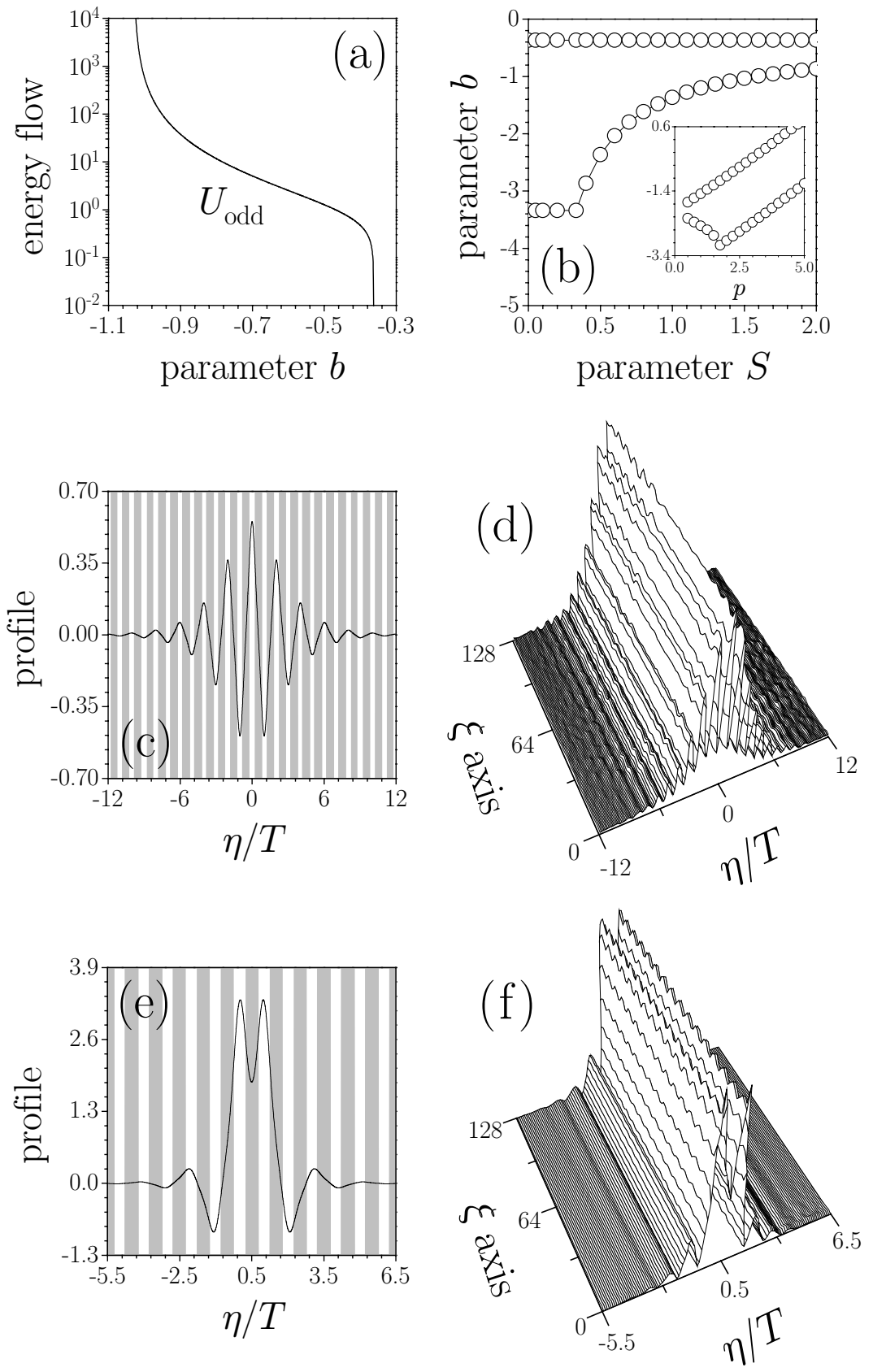

Figure 5. (a) Energy flow versus propagation constant for odd lattice soliton at $p = 3$, $S = 1.5$. (b) Lower and upper cut-offs for odd soliton versus saturation parameter at $p = 3$. Inset shows lower and upper cut-offs versus guiding parameter at $S = 0.5$. (c) Profile of odd lattice soliton at $p = 3$, $S = 1.5$, $b = -0.45$, and (d) its stable propagation in the presence of white noise. (e) Profile of even soliton at $p = 3$, $S = 0.4$, $b = -1.5$, and (f) its stable propagation in the presence of white noise. Noise variance in (d) and (f) $\sigma_{\mathrm{noise}}^2 = 0.02$. Defocusing medium $\sigma = 1$.